\documentclass[pre,aps]{revtex4}
\usepackage{graphicx,epsf}
\usepackage{amssymb}
\usepackage{amsmath}
\usepackage{latexsym}
\usepackage{amsthm,amsmath,amsfonts,amssymb,amsthm,amscd}
\usepackage{xurl}
\usepackage{xcolor}

\def\be{\begin{equation}}
\def\ee{\end{equation}}
\def\bq{\begin{eqnarray}}
\def\eq{\end{eqnarray}}
\def\beq{\begin{eqnarray*}}
\def\eeq{\end{eqnarray*}}

\makeatletter
\renewcommand{\p@subsection}{}
\makeatother

\begin{document}

\title{Comments on Weinstein's comments arXiv: 2509.09361 \& 2510.03793}

\author{Jean-Marc Ginoux}

\affiliation{Institut Jules Verne -- Toulon, \textcolor{blue}{jmginoux@orange.fr}\\
PhD in Applied Mathematics, Universit\'{e} de Toulon, \\
PhD in History of Science, Paris Sorbonne Universit\'{e}.}

\begin{abstract}
In 2024, after thirty years of research on this subject, I published a book entitled: \textit{Poincar\'{e}, Einstein and the discovery of special relativity. An end to the controversy} \cite{Ginoux2024}. In September 2025, Galina Weinstein published a review of this book entitled: \textit{Convergences and Divergences: Einstein Poincar\'{e} and Special Relativity} (arXiv:2509.09361) in which she harshly criticized my work in an unfair and error filled manner. The, she published a second comment (arXiv: 2510.03793) in which she added she added insults about me to all her mistakes, falsehoods and misleading criticisms. 
So I've decided to reply to her comments, in an academic way (as she would normally have done), in order to demonstrate that her allegedly ``novel way'' of reconstructing the history of the theory of special relativity is purely based on her own interpretation of the facts and not on the facts themselves. To this aim, I will follow the structure of each Weinstein's comment (arXiv: 2509.09361 \& 2510.03793) and I will highlight section by section all the erroneous things she has reported and repeated.
\end{abstract}

\maketitle

\section{Weinstein's comment arXiv: 2509.09361}

\subsection{Introduction of Weinstein's comment arXiv: 2509.09361}

In her introduction, Galina Weinstein claims to reconstruct ``in a novel way, the 1905 derivations of Einstein and Poincar\'{e}.'' Unfortunately, all the mathematical derivations she pretends to ``reconstruct'' at subsection 2.4. and in sections 3. have been already presented in my book \cite{Ginoux2024} in more detailed way at Chapter 6 (her subsection 2.4) and in Miller's contributions \cite{Miller1973,Miller1980,Miller1981,Miller2001}. Thus, her ``novel way'' dates back at least of 1973 and may be before.

\subsection{Between Convention and Innovation of Weinstein's comment arXiv: 2509.09361}

\subsubsection*{2.1 The Elephant in the Room}

In this section Galina Weinstein recalls:

\begin{quote}

``Ginoux also comments on Einstein's 1955 letter to Carl Seelig, in which Einstein professed ignorance of Poincar\'{e}s 1905 note [Poi05-1] and earlier Lorentz
papers. He characterizes this as ``surprising'' in light of Einstein's documented familiarity with contemporary literature, both in his published citations of the March 1905 quanta of light paper and in his role as a reviewer for the \textit{Beibl\"{a}tter zu den Annalen der Physik} [Gin]. While such claims bear on questions of influence rather than of strict priority, they invite broader historiographical reflection on the distinction between acquiring a mathematical toolkit and constructing a new conceptual architecture.''

\end{quote}

I still claim that it's very surprising that Einstein ignored the work of his predecessors in this article entitled ``On Electrodynamics of moving bodies'', while in his seven other articles published between 1901 and 1905 in \textit{Annalen der Physik} he cited all the most important work of his predecessors. Moreover, as a reviewer for the \textit{Beibl\"{a}tter zu den Annalen der Physik}, he was therefore well-versed in the scientific publication practices of his day.\\

To defend her point of view, Galina Weinstein cites Einstein's 1955 letter to Carl Seelig in which Einstein explains that: ``he knew Lorentz's 1895 work but neither Lorentz's later writings nor Poincar\'{e}'s related investigations, and that his 1905 work was ``in this sense'' independent.''\\

Here, the conflict of interest is obvious since the only witness on which these claims are based is Einstein himself. So, we must believe him according to Galina Weinstein because Einstein necessarily always tells the truth. This is unfortunately not the case and the biographies (see \textit{Albert Einstein Demystified}, Ginoux \cite{Ginoux2020}) I wrote on Einstein demonstrate this. Indeed, Einstein, like many others, lied to his wife, his children and also to his colleagues. So, why should we believe what he says about this article? Sorry but this is clearly not enough.

\newpage

Then, Galina Weinstein wrote:

\begin{quote}

``What was new, he emphasized, was recognizing that the Lorentz transformation applied beyond electrodynamics, reaching the general structure of space and time, and that Lorentz invariance was a universal constraint on physical theories.''

\end{quote}

In fact, the original Lorentz transformation of 1904 was uncomplete and so, was not invariant. In May 1905, one month before Einstein has submitted his article to \textit{Annalen der Physik}, Poincar\'{e} had already stated a complete transformation to which he gave the name of Lorentz and for which he proved its invariance. Moreover, it was not until 1910 that Einstein gave to the transformation (that he has established more than one month after Poincar\'{e}) the name of Lorentz. Thus, Einstein established on June 30, 1905 a transformation which is perfectly identical to that Poincar\'{e} stated at least one month before and which is not the original Lorentz transformation.\\

In the following of her subsection 2.1, Galina Weinstein makes reference to Gerald Holton and explains that:

\begin{quote}

``The omission of citations to either Poincar\'{e} or Lorentz's 1904 paper, he suggests, is best read in light of Einstein's normal practice of acknowledging sources he actively used; in the very same paper, Einstein twice names Lorentz when referring to the electron theory as presented in the 1895 Versuch, which he had read [Ein05, Hol60].''

\end{quote}

How could Holton know wether or not Einstein had actually used the articles that he quoted? Many searchers quote some papers they have not even read and many searchers don't quote for many reasons some papers they have read. This argument is simply unacceptable.\\

Then, Galina Weinstein writes that:

\begin{quote}

``Where Lorentz (and Poincar\'{e}) began from the transformations as a given, Einstein deduced them from two postulates - the relativity principle and the constancy of the speed of light - thus arriving by a distinct route [Hol60].''

\end{quote}

This sentence leads to a question concerning the development of the theory of special relativity.

\begin{center}
\textbf{How and why Lorentz was led to develop his transformation of 1904?}
\end{center}

He used the \textit{principle of relativity} formalized by Poincar\'{e} at the Saint Louis congress in 1904 and according to which ``the laws of physical phenomena must be the same for a stationary observer as for an observer carried along in a uniform motion of translation'' \cite{Poin1904} to show that Maxwell's equations of electromagnetism are invariant. Of course, Lorentz and Poincar\'{e} began from the transformation (without s) to prove its invariance according to the \textit{relativity principle}. This is confirmed by Lorentz himself who wrote in 1921 (nearly ten years after Poincar\'{e}'s death):

\begin{quote}
``I did not succeed in obtaining the exact invariance of the equations (\ldots).\\
I did not establish the principle of relativity as rigorously and universally true. Poincar\'{e}, on the contrary, obtained a perfect invariance
of the equations of electrodynamics, and he formulated the 'postulate of relativity', terms which he was the first to employ (\ldots) \cite{Lorentz1921}.''
\end{quote}

Then, Galina Weinstein explains:

\begin{quote}
``In Poincar\'{e}'s presentation, simultaneity, time measurement, and the operational meaning of coordinates remained within the conceptual boundaries
of the ether theory.''
\end{quote}

Here, Galina Weinstein refers to Poincar\'{e}'s article entitled ``La mesure du temps'' published in 1898 \cite{Poin1898}. Unfortunately, this contribution of Poincar\'{e} does not contain the word ether! Galian Weinstein discusses the question of ether in subsection 2.5 we analyze below.

\newpage

\subsubsection*{2.2 Between Bern and Paris: No Telegraph}

In this subsection Galina Weinstein writes:

\begin{quote}
``At that time, the 25-year-old patent examiner in Bern stood outside the scholarly correspondence networks through which such material typically circulated and had not yet met Lorentz. Surviving documentation records no communication from Lorentz to Einstein in this period, and Einstein's first known exchange with a leading academic - his correspondence with Max Planck - dates from roughly a year later.''
\end{quote}

This fairy tale reported from year to year by some historians of science is a pure fiction. As I have explained in my book \cite{Ginoux2024}, the links are actually quite numerous. First of all, Planck was associate editor of the journal \textit{Annalen der Physik}, in which Einstein published his first articles as early as 1901. Indeed, his famous article ``On the electrodynamics of moving bodies,'' considered the founding text of the special theory of relativity, was in fact his eighth publication in this journal. In addition, from 1905 onwards, Einstein wrote reviews of articles published in other international journals for the \textit{Be\"{i}bl\"{a}tter zu den Annalen der Physik}, i.e., the supplement to \textit{Annalen der Physik} (see Chap. 7 of my book). Moreover, according to Klein and Needell \cite{Klein}, Einstein reviewed a work by Planck published in 1906. An analysis of Einstein's correspondence shows that their first epistolary exchanges date back to the 6th July 1907. This is absolutely impossible, for several obvious reasons. Firstly, since Einstein worked for the supplement of \textit{Annalen der Physik}, he must have had contact with its editors, at least for signing his employment contract or for sending back his article reviews. Moreover, Einstein was still looking for an academic position at the university, as evidenced by this letter to his girlfriend Mileva Maric dated the 4th April 1901:

\begin{quote}
``Soon I will have honored all physicists from the North Sea to the southern tip of Italy with my offer!\cite{EinCorr1}.''
\end{quote}

It is therefore astonishing that Einstein did not send his application to Planck. Finally, in a letter from Einstein to his friend Maurice Solovine dated the 27th April 1906, we read:

\begin{quote}
``My papers are much appreciated and are giving rise to further investigations. Professor Planck (Berlin) has recently written to me about that \cite{EinCorr5}.''
\end{quote}

It is clear from this letter that Planck and Einstein had already corresponded prior to the 27th April 1906. Unfortunately, these letters have disappeared.\\

At the end of this subsection, Galina Weinstein writes the following sentence that is the leitmotif of her article:

\begin{quote}
``Poincar\'{e}'s synchronization and Einstein's synchronization may look similar at the procedural level, but they are embedded in fundamentally different
conceptual frameworks.''
\end{quote}

and that we could summarize as follows: \textbf{Same but Different}.

\subsubsection*{2.3 Priority Thread, in One Breath}

In this subsection Galina Weinstein writes:

\begin{quote}
``Ginoux's book adopts a formalist, sequence-oriented historiography, in which the systematic collation of equations, dates, and correspondence is used to reconstruct the relative timing and scope of contributions. On this basis, he attributes to Poincar\'{e}, by May-June 1905, a body of results encompassing the
corrections to Lorentz's 1904 formulas, the symmetric transformation form with $l = 1$, the group property, and the relativistic velocity-addition law as presented in the June 5 \textit{Comptes rendus} note [Gin]. For Ginoux, these achievements, combined with Poincar\'{e}'s articulation of the relativity principle, constitute the formal underpinnings of special relativity.''
\end{quote}

Let us recall here that this not Ginoux who considered that these achievements ``constitute the formal underpinnings of special relativity'' but Einstein himself. In fact, Einstein wrote in 1935 that he considered:

\begin{quote}
\centerline{``the Lorentz transformation [as] the real basis of the special relativity theory \cite{Einstein1935}.''}
\end{quote}

\newpage

At the end of this subsection, Galina Weinstein repeats again:

\begin{quote}
``Verbal correspondences, such as the parallel between the title of Einstein's paper and a phrase from Poincar\'{e}'s 1904 Saint Louis lecture [Poi04], are noted alongside recurrent juxtapositions of formulations that are often described as ``identical'' in substance, despite differences in expression.''
\end{quote}

As recalled in my book \cite{Ginoux2024}, in 1904 at Saint Louis, Poincar\'{e} concluded his oral presentation by this sentence:

\begin{quote}
``It is a question before all of endeavoring to obtain a more satisfactory \textit{theory of the electrodynamics of moving bodies} \cite{Poin1904}.''
\end{quote}

This expression is very interesting because it is exactly the title of Albert Einstein's article, ``Zur Elektrodynamik bewegter K\"{o}rper'' (On the electrodynamics of moving bodies), \textit{Annalen der Physik}, 17(10), 891-921, received the 30th June 1905 and published the 26th September 1905.\\

Thus, is there any difference? Absolutely none. This doesn't prove that Einstein has read Poincar\'{e}'s contribution of 1904. This is just a striking coincidence and there are many more.

\subsubsection*{2.4 The Ghost Prefactor}

This is probably the most surprising part of Weinstein's article. Concerning the following factor introduced by Einstein in his article:

\begin{equation}
\label{eq1}
\varphi(v) = a(v) \dfrac{1}{\sqrt{1 - \frac{v^2}{c^2}}}
\end{equation}

she wrote:

\begin{quote}
``Ginoux treats (1) as a purposeful nudge toward the Lorentz transformation rather than a neutral reparametrization..''
\end{quote}

First of all, it is not only Ginoux who ``treats (1) as \ldots'' but also Professor Arthur I. Miller. Then, it is unclear how and why Einstein introduced this factor as recalled by Miller who wrote in 1981 \cite{Miller1981}:

\begin{quote}
``How did Einstein know that he had to make the further substitutiona equals $a = \varphi(v)\sqrt{1 - \frac{v^2}{c^2}}$ in order to arrive at those space and time transformations in agreement with the postulates of relativity theory?''
\end{quote}

and he added:

\begin{quote}
``But why did Einstein make this replacement. It seems as if he knew beforehand the correct form of the set of relativistic transformations \cite{Miller1981}.''
\end{quote}

Unfortunately, Galina Weinstein seems to be unable to answer to these questions. At the end of this subsection Galina Weinstein makes a new mistake by writing:

\begin{quote}
``Using the final Lorentz transformation (38), after fixing $a(v)$ and $\varphi(v) = 1$, he obtained the relativistic addition law [Ein05].''
\end{quote}

Contrary to what has written Galina Weinstein, in his original paper, Einstein first used a long and tedious computation to obtain the relativistic addition law (see my the chapter 7 of my book \cite{Ginoux2024}). In Fig. \ref{fig1} below, I have reproduced a screen shot of Einstein's original paper \cite{Einstein1905} concerning his proof of the relativistic addition law

\begin{figure}[htbp]
\includegraphics[width=12cm,height=18cm]{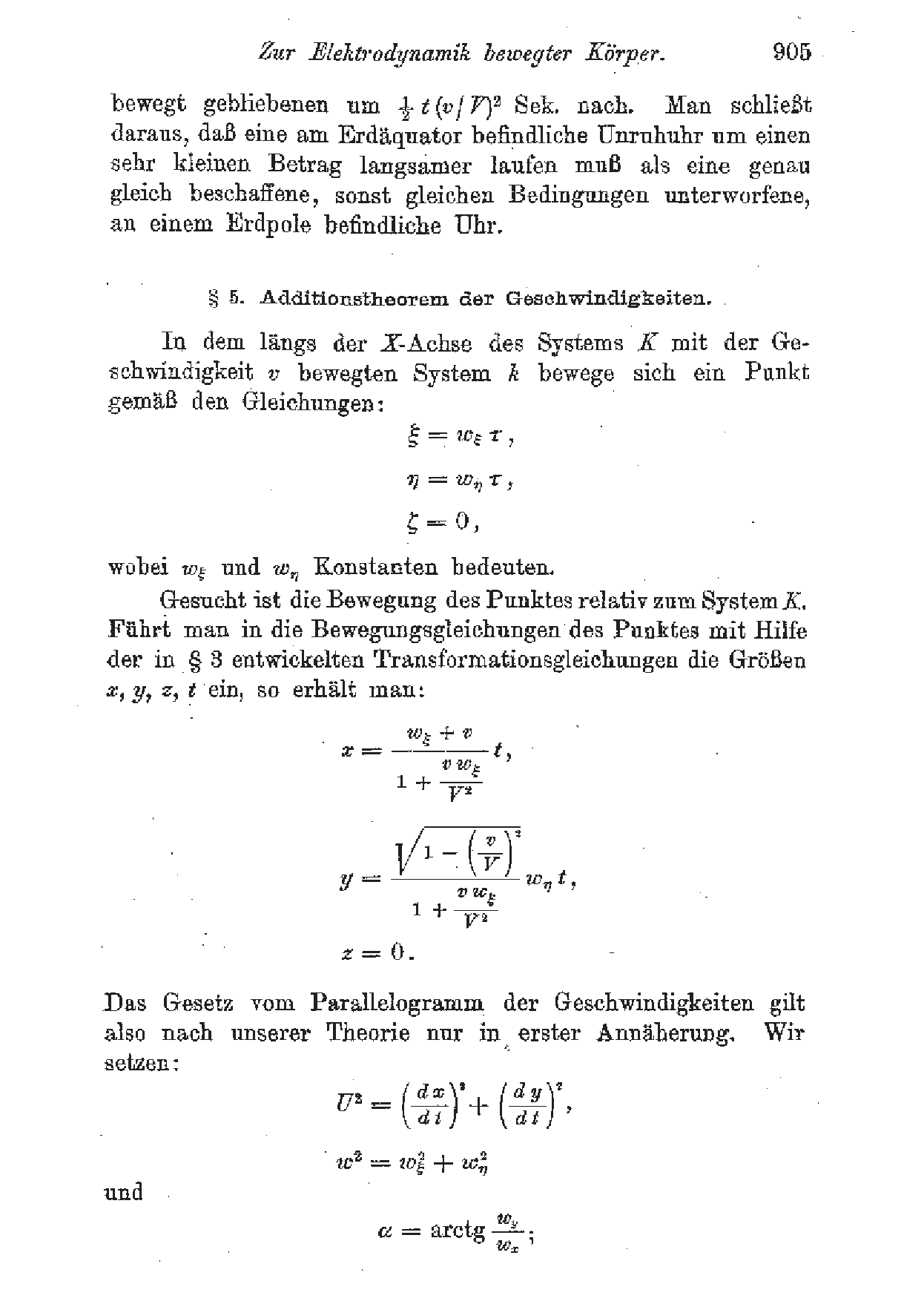}
\caption{Screen shot of Einstein's original paper}
\label{fig1}
\end{figure}

\newpage

As one can see in Fig. \ref{fig1}, contrary to what claims Galina Weinstein, this is not by using the final Lorentz transformation that Einstein first obtained the relativistic addition law.\\

At the end of this subsection, Galina Weinstein explains:

\begin{quote}
``Ginoux suggests that the paucity of citations in 1905 indicates dependence on Lorentz / Poincar\'{e}. Regardless of historical editorial practice, that claim is orthogonal to the logic of the derivation.''
\end{quote}

Here again, Galina Weinstein is in the interpretation of what I have written. I have just explained in my book that it's important to demonstrate that the argument too often put forward by some historians of science that ``it wasn't customary to cite one's sources in 1905''(as it seems to be the case for Galina Weinstein here) doesn't hold water. In fact, three months before the publication of his article ``On the electrodynamics of moving bodies'', Einstein published an article entitled ``On a heuristic point of view concerning the production and transformation of light'' \cite{Einstein1905a} on March 18, 1905, also in \textit{Annalen der Physik}. In this work on the explanation of the photoelectric effect, for which he was awarded the Nobel Prize in Physics in 1922 (for the year 1921), Einstein made no fewer than seven quotations.\\

\textbf{So why does he quote his peers in March and not in June?} Once again, this fact raises questions.

\subsubsection*{2.5 Poincar\'{e}'s Ether vs. Einstein's Ether}

In this subsection Galina Weinstein discussed the existence or non-existence of ether which would be a decisive step in the development of the theory of special relativity theory. Here again, this is not the good question to be addressed. Of course Poincar\'{e} has kept the \textit{luminiferous ether} as many other scientists at that time, i.e. like Lorentz and Planck and before them James Clerk Maxwell. At that time, it was commonly admitted that light was propagating in a medium called \textit{luminiferous ether}. Contrary to what claims Galina Weinstein, Einstein did not abolish ether. In his article, Einstein \cite{Einstein1905} exactly said: ``Die Einf\"{u}hrung eines ``Licht\"{a}thers'' wird sich insofern als \"{u}berfl\"{u}ssig erweisen'' (The introduction of a ``light ether'' will prove superfluous in this respect). The word ``\"{u}berfl\"{u}ssig (superfluous)'' does not mean that he abolished ether but that he didn't need to use it as it was nearly the case. In fact as highlighted in my book, Lorentz and Poincar\'{e} were working on a problem of contraction of electron moving at velocity close to that of light. Thus, they tried to explain such a contraction by means of force and more particularly the Lorentz electromagnetic force. Thus, their approach was \textit{dynamic} (from the Greek \textit{dyn} which means force). The title of Poincar\'{e}'s two main contributions entitled ``Sur la dynamique de l'\'{e}lectron'' (On the dynamics of the electron) \cite{Poin1905, Poin1906} is enough to prove that. The reason why Poincar\'{e} kept the \textit{luminiferous ether} is based on his \textit{dynamic approach} as highlighted by this sentence he wrote in 1900:

\begin{quote}
``If we did not wish to change the whole of the science of mechanics, we should have to introduce the ether, in order that the action which matter apparently undergoes should be counterbalanced by the reaction of matter on something. \cite{Poin1900a}.''
\end{quote}

Poincar\'{e} justified the existence of a \textit{luminiferous ether} as a convenient hypothesis, explaining that without it, Newton's third law - the action-reaction principle - would no longer be respected. Thus, his \textit{dynamic} approach consisted in extending Newton's classical mechanics to Maxwell's electromagnetism in what he called a \textit{New Mechanics} (\textit{M\'{e}canique Nouvelle}). By completing the latest transformation provided by Lorentz in 1904, Poincar\'{e} proved that the resulting transformation he had thus obtained formed a group of invariance of the Dynamics.\\

Einstein's article \cite{Einstein1905} shares into two main sections entitled \textit{kinematic} and \textit{dynamic} part. As recalled by the title of the first part, his approach is, first of all, \textit{kinematic}. Thus, his aim is completely different from that of Lorentz and Poincar\'{e}. At first, he is not interested in the contraction of an electron moving at a velocity close to that of light. The problem to which he is faced is the Doppler-Fizeau effect. This phenomenon can easily be observed in our everyday lives when an emergency vehicle passes by. When the vehicle approaches, the sound produced by its siren seems higher-pitched, whereas it seems lower-pitched when it moves away. The question for Einstein was then to know what would happen if we replace the acoustic or mechanical wave, i.e. the siren, with an electromagnetic wave, i.e. light. To this aim, Einstein used this famous metaphor: if a person travels at a speed close to that of light, will he be able to see his face in a mirror placed in the direction of his travel? Following this idea, Einstein analyzes motion independently of the causes (forces) that produce it. More precisely, Einstein studied the propagation of a light signal from one frame of reference to another. However, as recalled above, normally Einstein should have kept the \textit{luminiferous ether} since at that time, it was commonly admitted that light propagates in such a medium. But he didn't say anything about the medium in which his light signal propagated. It's only in the second part of his article that he considers an empty space, without giving any definition. By using his \textit{kinematic} approach, he was able to establish a transformation which is exactly that given by Poincar\'{e} in May 1905 and published by Poincar\'{e} \cite{Poin1905} on June 5$^{th}$ 1905 in the \textit{Comptes Rendus}, thus at least three weeks before Einstein had submitted his article. \\

At the end of this subsection, Galina Weinstein explains:

\begin{quote}
``When Einstein did speak of an ether, it was in a sense that differed fundamentally from Lorentz's construct.''
\end{quote}

To support it, Galina Weinstein repeats the sentence uttered at the time by John Stachel, whom she thanks at the end of her article, as if repeating an argument could give it a truth value:

\begin{quote}
``the ether he reintroduced differed fundamentally from the ether he had banished [Sta-01].''
\end{quote}

But to defend the indefensible, Galina Weinstein will now provide us with the following weighty argument.

\begin{quote}
``In a 1916 letter to Lorentz, Einstein himself, half-diplomatically and half-seriously, had remarked that general relativity was ``closer to the ether hypothesis'' than special relativity [CPAE8], Doc. 222. The remark was a gesture of respect: Lorentz still clung to ether, and Einstein, who revered him, framed his own theory in language that Lorentz could recognize. The Leiden lecture was thus homage as much as physics - \textbf{a theatrical bow to Lorentz}, even as the stage and script had already shifted.''
\end{quote}

Thus, according to Galina Weinstein, when Einstein reintroduced the ether in a lecture given at Leiden, this is just ``a theatrical bow to Lorentz'' and we should not consider that he really reintroduced the \textit{luminiferous ether}. Let us recall below the conclusion of Einstein's lecture in Leiden \cite{Einstein1921} which has been subtly omitted by Galina Weinstein:

\begin{quote}
``Recapitulating, we may say that according to the \textit{general theory of relativity} space is endowed with physical qualities; in this sense, therefore, \textbf{\textit{there exists an ether}}. According to the \textit{general theory of relativity} \textbf{\textit{space without ether is unthinkable}}; for in such space there not only would be no propagation of light, but also no possibility of existence for standards of space and time (measuring-rods and clocks), nor therefore any space-time intervals in the physical sense. But this \textbf{\textit{ether}} may not be thought of as endowed with the quality characteristic of ponderable media, as consisting of parts which may be tracked through time. The idea of motion may not be applied to it \cite{Einstein1921}.''
\end{quote}

This quotation is very interesting because it proves that in his 1905 paper, Einstein \cite{Einstein1905} should not have considered the \textit{luminiferous ether} as superfluous because, as he wrote in his conclusion above, in a space without ether, propagation of light would be impossible. Moreover, when Einstein has decided to generalize the special relativity theory, i.e. to analyze the gravitational force able to bent light, he was obliged to reintroduce an ether as Poincar\'{e} did before him to analyze in a \textit{dynamic} approach the contraction of an electron moving at velocity close to that of light. Thus, the argument of Galina Weinstein according to which the Leiden lecture was ``a theatrical bow to Lorentz'' is simply irrelevant. To confirm that Einstein had really introduced ether in the framework of general relativity theory, let us recall that in January 1920 Einstein had written a remarkable article for the journal \textit{Nature}, which was never published and which he probably used for his lecture in Leiden. In his conclusion, Einstein wrote:

\begin{quote}
``Again, empty space seems to be endowed with physical properties, that is, not physically empty as it appeared in the \textit{special theory of relativity}. Therefore, \textbf{one can say the ether has been resurrected in the theory of general relativity}, even though in a (newer) more sublime form. The ether of the \textit{general theory of relativity} differs from the one in old optics by not being a substance in the sense of mechanics. Not even the concept of motion can be applied to it \cite{EinCorr7}.''
\end{quote}

\subsubsection*{2.6 The Light Postulate}

In this subsection Galina Weinstein explains:

\begin{quote}
``Ginoux regards Einstein's 1905 light postulate as curious, since it appears to place the invariance of $c$ at the origin of the relativity principle, rather than the other way around.''
\end{quote}

I still claim that Einstein's 1905 light postulate is curious, since in my opinion it should have been like for Poincar\'{e} a consequence of the \textit{relativity principle} and not the contrary. From the \textit{relativity principle}, one can deduce the classical addition-law which is no more valid for light as proven by Michelson-Morley experiments. So, Einstein is perfectly right when he explains that:

\begin{quote}
``We will raise this conjecture (the purport of which will hereafter be called the \textit{Principle of Relativity}) to the status of a postulate, and also introduce another postulate, which is only apparently \textbf{irreconcilable} with the former, namely, that light is always propagated in empty space with a definite velocity $c$ which is independent of the state of motion of the emitting body. \cite{Einstein1905}.''
\end{quote}

However, in my opinion, the invariance of velocity of light is not a postulate \textbf{irreconcilable} with the \textit{principle of relativity} but a consequence of this principle. That's the reason why I invoked the works of Pr. Jean-Marc L\'{e}vy-Leblond \cite{LevyLeblond1976}.\\

In this subsection Galina Weinstein writes:

\begin{quote}
``The identification of $c$ as the limiting speed remains an essential, physically motivated step - one that Einstein's 1905 formulation incorporates from the outset with full conceptual economy.''
\end{quote}

Let us recall that at Saint Louis, in September 1904, Poincar\'{e} had already stated the principle of invariance of the velocity of light:

\begin{quote}
``From all these results, if they were confirmed, would arise an entirely new mechanics, which would be, above all, characterized by this fact, that no velocity could surpass that of light \cite{Poin1904}.''
\end{quote}

In May 1905, in one of the letters written to Lorentz (see Fig. 2), Poincar\'{e} explained that in his demonstration, aimed at completing the Lorentz transformation, he has chosen the units such that of light $c = 1$. Obviously, if Poincar\'{e} posed $c=1$, it is because he considered the velocity of light was the same in all reference frames.

\subsection{Einstein's and Poincar\'{e}'s Derivations of Weinstein's comment arXiv: 2509.09361}
\subsubsection*{3.1 Poincar\'{e}'s May 1905 Letters to Lorentz}

In this subsection, Galina Weinstein reproduces parts of what I have published in Chapters 5 \& 6 of my book \cite{Ginoux2024}. Then, she recalls that by applying successively two transformations, Poincar\'{e} stated the following formula:

\begin{equation}
\label{eq2}
\varepsilon'' = \dfrac{\varepsilon + \varepsilon'}{1 + \varepsilon \varepsilon' }
\end{equation}

Then, Galina Weinstein explains:

\begin{quote}
``Why is equation (26) (Eq. (2) above) not the velocity addition law? The symbols $\varepsilon$, $\varepsilon'$, $\varepsilon''$ are group parameters labeling Lorentz transformations (essentially, dimensionless rapidities). Equation (26) is therefore the group composition law for successive
boosts, not a physical law of how material particle velocities add.''
\end{quote}

Unfortunately, this argument is again irrelevant because Poincar\'{e} wrote in his letters to Lorentz reproduced in Figs. 2 \& 3 below.

\begin{quote}
\centerline{``Let $-\varepsilon$ be the speed of translation with that of light taken as unity.''}
\end{quote}

\begin{figure}[htbp]
\includegraphics[width=15.42cm,height=14.68cm]{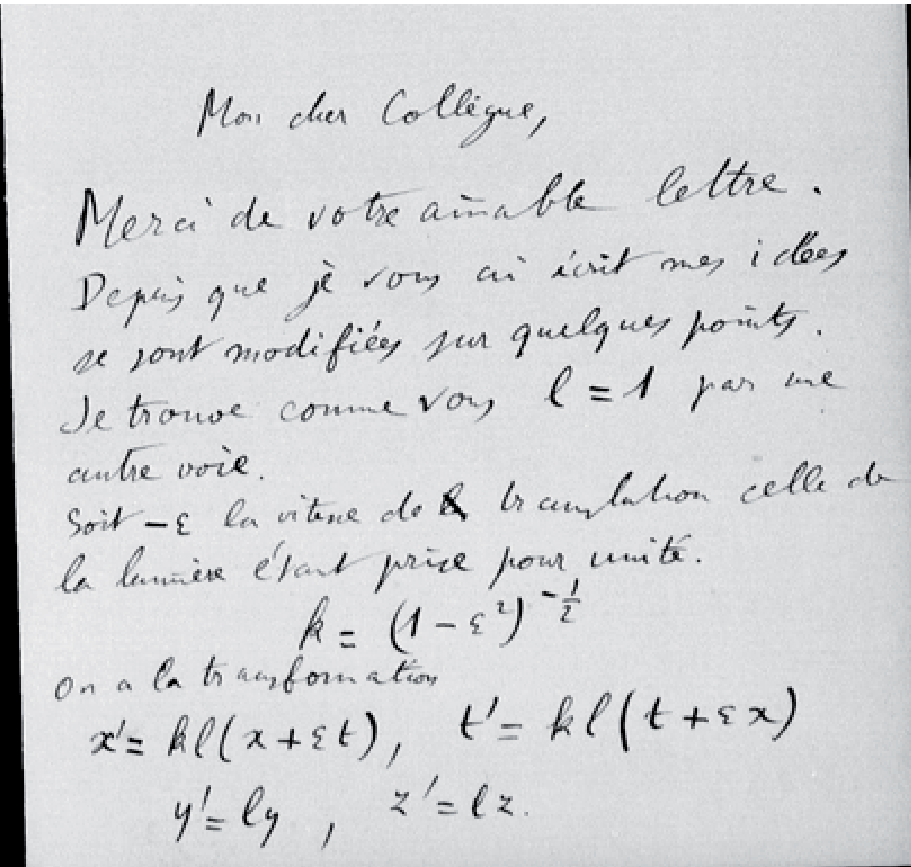}
\caption{Poincar\'{e}'s May 1905 original letter to Lorentz, by courtesy of Noord-Hollands Archief, [Fonds Poincar\'{e}], NHA-9423.}
\label{fig2}
\end{figure}

\begin{figure}[htbp]
\includegraphics[width=15.94cm,height=16.77cm]{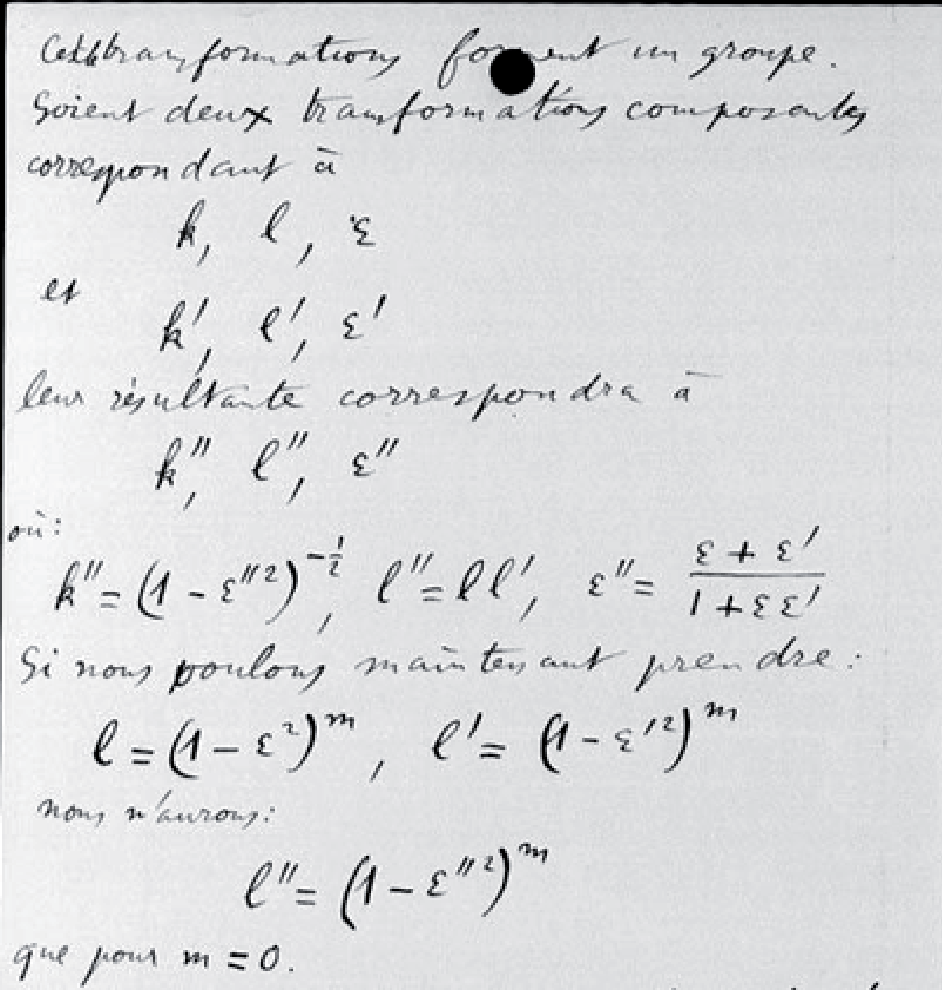}
\caption{Poincar\'{e}'s May 1905 original letter to Lorentz, by courtesy of Noord-Hollands Archief, [Fonds Poincar\'{e}], NHA-9423}
\label{fig3}
\end{figure}

It follows that the symbols $\varepsilon$, $\varepsilon'$, $\varepsilon''$ are not group parameters but ``speed of translation'' according to Poincar\'{e} himself. Thus, the formula established by Poincar\'{e} in his May 1905 letters to Lorentz (see Figs. \ref{fig2} \& \ref{fig3}) is indeed the \textit{relativistic velocity addition law}.

\subsubsection*{3.2 Poincar\'{e}'s 1906 Derivation in the \textit{Rendiconti} paper}

In this subsection, Galina Weinstein reproduces parts of what I have published in Chapters 5 \& 6 of my book \cite{Ginoux2024}. From her computations she obtains the following equation:

\begin{equation}
\label{eq3}
\xi' = \dfrac{\xi + \varepsilon}{1 + \varepsilon \xi }
\end{equation}

Then, she explains in the previous subsection that:

\begin{quote}
``The actual velocity transformations (36) (Eq. (3) above) require differentiating the Lorentz transformation (24) as Poincar\'{e} himself later derived in the \textit{Rendiconti di Palermo} memoir.''
\end{quote}

In the subsection 3.2 she confirms that her equation (36) (Eq. (3) above) is indeed the \textit{relativistic velocity addition law} because in his \textit{Rendiconti di Palermo} memoir Poincar\'{e} has used a \textit{differential operator} to derive it. This is simply incredible! The first argument of Galina Weinstein presented in her subsection 3.1 was that Poincar\'{e} used group symbols. But here, $\xi$ and $\varepsilon$ are still dimensionless group symbols. So, according to her own argument, her equation (36) (Eq. (3) above) should not be considered as the \textit{relativistic velocity addition law}.\\

Could Galina Weinstein explain us the difference between her equations (36) and (26) (Eq. (3) and Eq. (2) above).\\

There are indeed two different ways to derive the \textit{relativistic velocity addition law}. The first consists in proving that the Lorentz transformation is invariant. This is what Poincar\'{e} has done in his May 1905 letters to Lorentz (see Figs. 2 \& 3). The second consists in using the differential operator. This is what Poincar\'{e} has done in his \textit{Rendiconti di Palermo} memoir but not Einstein in the first part of his article \cite{Einstein1905} (see Fig. 1). Let us notice that if the last method does indeed provide the \textit{relativistic velocity addition law}, it does not prove the invariance of the Lorentz transformation and therefore does not allow to demonstrate that it forms an invariance group of the Dynamics. Concerning Einstein, let us notice that he didn't prove that the Lorentz transformation forms an invariance group of the Dynamics. He wrote in his paper:

\begin{quote}
\centerline{``\ldots such parallel transformations - necessarily - form a group. \cite{Einstein1905}''}
\end{quote}

Moreover, this sentence without any proof is very surprising since at that time, Einstein had never spoken about the group theory neither in his publications nor in his correspondence while Poincar\'{e} had already been in contact several times with Lazarus Fusch and Felix Klein and had already published many papers on this subject. May be another striking coincidence for Einstein?

\newpage

\subsubsection*{3.3 Einstein's 1905 Derivation of The velocity and Charge Density Transformation}

In this subsection, Galina Weinstein presents Einstein's long and tedious computation of the velocity and charge density already presented in the chapter 7 of my book \cite{Ginoux2024}. This is also very important to recall that according to Einstein and so to Galina Weinstein, Einstein had never been aware of Lorentz 1904 paper \cite{Lorentz1904} in which he gave some erroneous expressions of the velocity and charge density. This has been confirmed by Lorentz in 1921. He wrote:

\begin{quote}
``Poincar\'{e}, on the contrary, obtained a perfect invariance of the equations of electrodynamics, and he formulated the \textit{postulate of relativity},
terms which he was the first to employ. Indeed, stating from the point of view that I had missed, he found the formula ([of the velocity and charge density]). Let us add that by correcting the imperfections of my work he never reproached me for them \cite{Lorentz1921}.''
\end{quote}

How could Einstein know that he had succeeded to obtain the correct expressions for the velocity and density of charge without reading Lorentz 1904 paper? Another coincidence probably? Let us notice that Einstein's expressions of the velocity and density of charge are the same as those obtained in May-June by Poincar\'{e}, as highlighted in the chapter 7 of my book \cite{Ginoux2024},

\subsubsection*{3.4 The Addition Law that Made the Difference}

In this subsection, Galina Weinstein explains that:

\begin{quote}
``Einstein did not, and in 1905 could not, adopt this purely formal route [group theory], since his method was heuristic in character.''
\end{quote}

This surprising to read that Einstein has the right to use a heuristic method to obtain the \textit{relativistic velocity addition law} while Poincar\'{e} cannot use such a method to keep the \textit{luminiferous ether}.

\subsection{In the Shadow of Light Beams of Weinstein's comment arXiv: 2509.09361}

\subsubsection*{4.1 The Relativity of Recognition}

In this subsection, Galina Weinstein recalls that:

\begin{quote}
``Ginoux's examination of the Nobel dossiers casts the episode less as a mystery of merit than as a study in how institutions quietly decide what will be
remembered. Despite repeated nominations, Poincar\'{e}'s case never advanced. The committee preferred tangible experiment to abstract theory, mistrusted
mathematical style in physics, and managed national sensibilities with a cautious hand. The reports are respectful, sometimes admiring, but never decisive.
His work is acknowledged in passing, yet consistently absorbed into Lorentz's theory or filed under collective progress. By the time of his death in 1912, the
record had settled into a pattern of praise without credit - an archive of tributes that effectively erased their subject.''
\end{quote}

Galina Weinstein has just omitted again two crucial points. The first is that starting from 1910, Einstein gave to the transformation he obtained in 1905 the name of Lorentz. Einstein wrote:

\begin{quote}
``These transformation equations were introduced in a very successful manner into electro-dynamics by M. Lorentz. We'll refer to them as the Lorentz transformation \cite{Einstein1910}.''
\end{quote}

Why didn't Einstein do it before? Probably another coincidence. The most probable reason is that if he had done that in 1905 it will have proven that he had read Poincar\'{e}. The second crucial point is that in 1912 one discovers that Wien nominated Lorentz and Einstein for their common discovery of the special theory of relativity. Lorentz's attitude is inexplicable to me. In May 1905, he writes to Poincar\'{e} to ask him his help concerning the transformation he had obtained gropingly \cite{Lorentz1921}, Poincar\'{e} send him back the complete transformation and the correct expressions for the velocity and density of charge. And a few years later instead of supporting him for the Nobel Prize of Physics (as Poincar\'{e} did for him in 1901) he applies for a second Nobel Prize with Einstein. This is very surprising even shocking. But there are probably some unknown reasons for this turnaround.

\subsubsection*{4.2 \textit{La probl\'{e}matique}: Why Did Poincar\'{e} Not Claim Authorship of Special Relativity?}

In this subsection, Galina Weinstein explains that:

\begin{quote}
``And yet, a few years later, in his 1912 London lecture, he was speaking the idiom of Minkowski space and edging toward Einstein's own perspective (\ldots)
One even hears the unmistakable echo of Minkowski in his remark that the fourth coordinate is best taken as $\sqrt{-1}$.''
\end{quote}

This is completely untruth. In his \textit{Rendiconti di Palermo} memoir \cite{Poin1906}, Poincar\'{e} introduced two years before Minkowski \cite{Minkowski1908} the concept of four-vector as highlighted by the page 168 of his famous memoir (see Fig. \ref{fig4} below).

\begin{figure}[htbp]
\includegraphics[width=15.24cm,height=14.59cm]{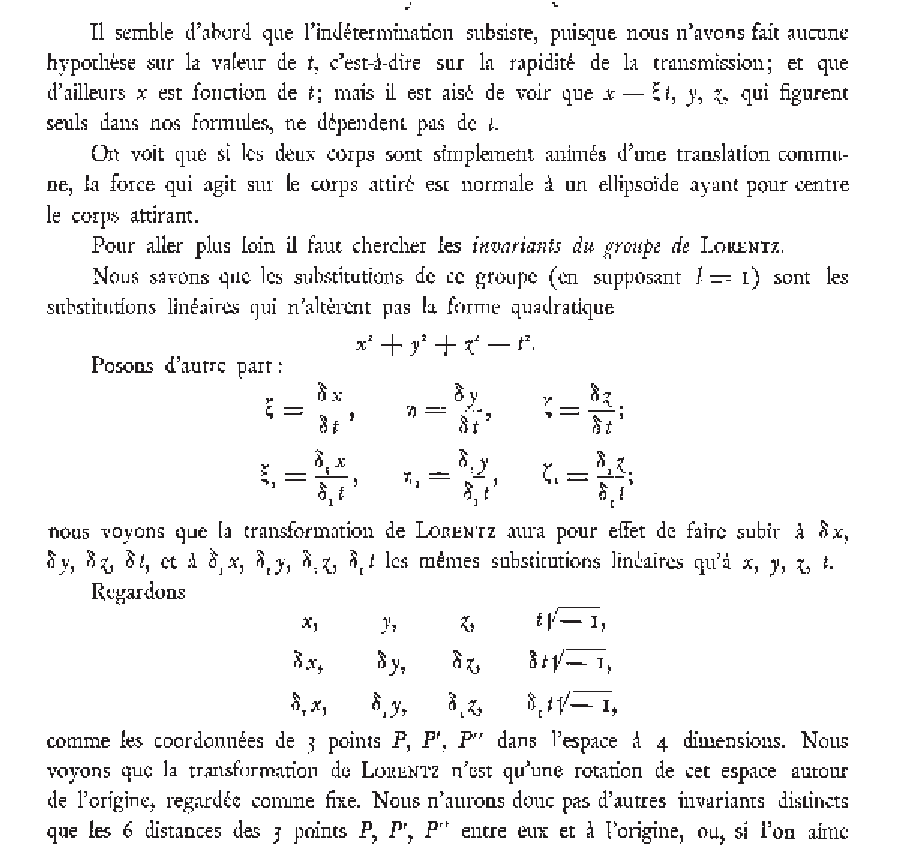}
\caption{Page 168 of Poincar\'{e}'s \textit{Rendiconti di Palermo} memoir \cite{Poin1906}.}
\label{fig4}
\end{figure}

Then, Galina Weinstein explains that:

\begin{quote}
``The irony is hard to miss: Ginoux asks why Poincar\'{e} never claimed authorship of relativity [Gin], yet Poincar\'{e}'s own words, in the last public address of his life, sound less like a claimant and more like a convert. The revolution, it seems, was already underway - only the naming rights remained unspoken.''
\end{quote}

This is again false. Poincar\'{e} gave his last lecture at the \'{E}cole Sup\'{e}rieur des Postes et T\'{e}l\'{e}graphes today SupTelecom Paris in July 1912 a few days before his death as confirmed by the subtitle \cite{Poin1912}.

\subsection{Conclusion: \textit{Saving Private Einstein}}

It is very surprising to observe for decades how some historians of science are able not to reconstruct but really to rewrite the development of the theory of special relativity by inventing imaginary facts or by interpreting real facts in a incredible manner as it is the case for the article of Galina Weinstein. First of all she explains that she will ``reconstruct in a novel way the 1905 derivations of Einstein and Poincar\'{e}'' although she had only recopied some results already published in my book \cite{Ginoux2024} and in the contributions of Pr. Miller \cite{Miller1973,Miller1980, Miller1981, Miller2001}. Thus, her ``novel way'' dates back at least of 1973 and may be before. Her aim is clearly to show that even if Poincar\'{e} has published the main results concerning the theory of special relativity (complete Lorentz transformation and its invariance while using group theory, relativistic velocity addition law, \ldots), Einstein must be considered as the unique father of this theory.\\
Then, Galina Weinstein uses different types of arguments to support her point of view. The first one is the ``truth''. As an example, Einstein claimed several times that he had never read Poincar\'{e}'s note and memoir, so it must be true and we must believe in his words even if he has lied in several circumstances to his wives, children and colleagues. If it is not enough, she invokes Gerald Holton according to whom Einstein only quoted papers he had read.\\
The second type of argument is ``the selection of facts'' which thus makes it possible to rewrite a history which corresponds to the thesis that one wishes to defend. As an example Galina Weinstein omitted to recall that the original Lorentz transformation of 1904 was uncomplete and that Lorentz contacted Poincar\'{e} in May 1905 to ask him to solve this problem.\\
The third type of argument is more dangerous but very classical: ``the falsehoods''. As an example, she explains that in his contribution entitled ``La mesure du temps'' Poincar\'{e} \cite{Poin1898} already used the concept of \textit{luminiferous ether} although even the expression does not appear in this paper as it is easy to verify. Then, she tries to make us believe that Einstein who was working for \textit{Annalen der Physik} since 1901 and who has published before 1905 eight articles in this revue had never been in contact with one of the editor in chief Max Planck.\\
Unfortunately for Einstein's defenders, the date of the publication of Poincar\'{e}'s contributions as well as that of Einstein is very well-known and well-documented: on June 5$^{th}$ 1905, Poincar\'{e} published his note \cite{Poin1905} and on June 30$^{th}$ 1905, Einstein submitted his article to \textit{Annalen der Physik} \cite{Einstein1905} and so, it is obvious that Einstein 1905 famous paper is therefore at least several weeks later.\\
Here appears a new kind of argument: ``same but different''. Poincar\'{e} and Einstein's results seem to be the same but they are different. Why? This is not crystal clear. Thus, in her article Galina Weinstein made use of these four kinds of arguments to support her point of view which can be summarized as follows. First, she is obliged to admit that Einstein's famous paper is subsequent to Poincar\'{e}'s contributions. So, she explains that contrary to Einstein, Poincar\'{e} kept the \textit{luminiferous ether} and this fact precludes to consider that he could have laid the foundations of the theory of special relativity. However, Galina Weintein is also obliged to recognize that during his Leiden lecture in 1920 before Lorentz, Einstein claimed that, in the framework of the theory of general relativity, ``there exists an ether''. Faced to such Einstein's inconsistency (this is not the only one), she explained that when Einstein reintroduced the ether in a lecture given at Leiden, this is just ``a theatrical bow to Lorentz'' and we should not consider that he really reintroduced the \textit{luminiferous ether}. Such an argument is simply irrelevant. But the most incredible is her argument to explain that Poincar\'{e} cannot have derived the \textit{relativistic velocity addition law} while using the invariance of group because the symbols involved are those of group parameters. In her next subsection she recognizes that Poincar\'{e} has finally derived the \textit{relativistic velocity addition law} while using a \textit{differential operator} applied to a transformation (that Poincar\'{e} has completed and to which he has given the name of Lorentz) involving the same group parameters symbols. Curiously, Galina Weinstein made no comment on the fact that Einstein provided no proof that the Lorentz transformation forms a group of invariance of the Dynamics. She also omitted to recall that Lorentz expressions for the velocity and charge density were erroneous. A fact that Einstein was supposed to ignore since he was (according to him) unaware of Lorenz 1904 publication. In her article, Galina Weinstein also omitted to recall that Lorentz and Einstein applied in 1912 for the Nobel Prize of Physics for their common discovery of the theory of special relativity. A fact that raises questions. Her last subsection is also for Galina Weinstein the occasion to present a fake news by claiming that Poincar\'{e} has made use of Minkowski's fourth-vector although Poincar\'{e} introduced it two years before him in his \textit{Rendiconti di Palermo} memoir \cite{Poin1906}.\\

It is very regrettable that some historians of science are capable of using such methods to defend an indefensible point of view which does not stand up to analysis of the facts. Let us recall to them Poincar\'{e}'s words:

\begin{quote}
``Thinking must never submit itself, neither to a dogma, nor to a party, nor to a passion, nor to an interest, nor to a preconceived idea, nor to whatever it may be, if not to facts themselves, because, for it, to submit would be to cease to be \cite{Poin1909}. ''
\end{quote}

\newpage

\section{Weinstein's comment arXiv: 2510.03793}

\subsection{Introduction of Weinstein's comment arXiv: 2510.03793}

In her introduction, Weinstein presents the main argument that she will uses in her answer: ``the same is different''. She explains

\begin{quote}
``That is, the difference between \textit{formal structures}--equations, group properties, calculational devices--that were indeed available to Lorentz and
Poincar\'{e} by mid-1905 and the \textit{conceptual framework} inaugurated by Einstein in June 1905, in which simultaneity is practically defined, the ether is rendered superfluous, and the Lorentz transformation is derived from two coequal postulates.''
\end{quote}

Thus, according to Mrs. Weinstein, the difference between the transformation that Poincar\'{e} sent to Lorentz in his letters dating from May 1905 and the transformation proposed by Einstein \cite{Einstein1905} in his famous paper submitted on June 30$^{th}$ to \textit{Annalen der Physik} is that the former is based on ``\textit{formal structures}--equations and group properties'' while the latter is based on ``\textit{conceptual framework}''. Just two questions to which I would like her to answer:

\begin{itemize}
\item[\textbf{Q1:}] \textbf{What are the mathematical differences between the two transformations?}
\item[\textbf{Q2:}] \textbf{Why this transformation is nowadays called Poincar\'{e} group?}
\end{itemize}

\subsection{Reply to Ginoux's \textit{Ad Hominem} Critique of Weinstein's comment arXiv: 2510.03793}

In this section, Mrs. Weinstein recalls Einstein's 1955 letter to Carl Seelig, in which Einstein professed ignorance of Poincar\'{e}s 1905 note \cite{Poin1905} and earlier Lorentz papers except that of 1895. I have written that I could give any credit to Einstein's claim since he has lied to his wives, children and colleagues during all his life. To this Mrs. Weinstein replies:

\begin{quote}
``This reasoning collapses private life into a wholesale claim of intellectual dishonesty. Methodologically, that is untenable; the personal failings of a scientist cannot be marshalled as evidence against their scholarly testimony. To reduce the question of influence in the genesis of special relativity to judgments about moral character is to leave the historical method for insinuation.''
\end{quote}

Just a unique question:

\begin{itemize}
\item[\textbf{Q3:}] \textbf{How can you prove that what Einstein says is true?}
\end{itemize}

In fact, Mrs. Weinstein can prove neither this claim is true nor false. Thus, since Mrs. Weinstein pretends to follow a ``\textit{a strictly academic manner}'', she should at least challenge this affirmation but instead of doing this, she chooses to believe what Einstein says without any doubt. Let suppose that we should follow  such an argument.\\

Professor Wilhelm R\"{o}ntgen, who discovered the famous X-rays and won the first Nobel Prize in physics in 1901, received Einstein's manuscript on the electrodynamics of moving bodies. R\"{o}ntgen instructed one of his assistants, the Russian physicist Abram Joffe, a future member of the USSR
Academy of Sciences, to examine the article. However, the latter will report, in his \textit{Souvenirs d'Albert Einstein}, that the original of the article, destroyed since, was jointly signed Einstein-Maric. So Einstein's wife participated in the writing of the article?\\

Thus, Mrs. Weinstein should admit without any doubt that Joffe told the truth and and that Einstein's original article on special relativity theory was co-authored with his first wife. But, I am pretty sure that Mrs. Weinstein will explain that in this case even if we cannot prove that this claim is true or false we must consider that it must be false. At the end of her section in which she quotes many historians of science who believe the same, she explains:

\begin{quote}
``The argument requires engagement with textual evidence, not dismissal by analogy.''
\end{quote}

Textual evidence? So the necessary and sufficient condition for something to be true is simply to have been written by Einstein? Clearly, this is not enough!
To begin to believe a statement, it is necessary to have at least a second source that corroborates the first. This is an academic way of proceeding. \\

Let us take another example of this ``textual evidence'' so important to Mrs. Weinstein. After Poincar\'{e}'s untimely death on the 17$^{th}$ July, 1912, an unanimous tributes was paid to him by most of his colleagues and friends. Among them, Lorentz wrote an analysis of Poincar\'{e}'s work entitled ``Sur la dynamique de l'\'{e}lectron'' (On the dynamics of the electron):

\begin{quote}
``These were the considerations published by me in 1904 which gave place to Poincar\'{e} to write his paper on the dynamics of the electron, in which he attached my name to the transformation to which I will come to speak. I must notice on this subject that the same transformation was already present in an article of Mr. Voigt published in 1887, and that I did not draw from this artifice all the possible parts. Indeed, for some of the physical quantities which enter the formulas, I did not indicate the transformation which suits best. \textbf{\textit{That was done by Poincar\'{e} and then by Mr. Einstein}} and Minkowski \cite{Lorentz1921}.''
\end{quote}

Then, he added:

\begin{quote}
``Poincar\'{e}, on the contrary, obtained a perfect invariance of the equations of electrodynamics, and he formulated the ``postulate of relativity'', terms which he was the first to employ.''
\end{quote}

At first glance, Lorentz seems to credit Poincar\'{e} with the foundations of the \textit{special theory of relativity}, if not the theory itself. And yet, a few months later, the same Lorentz wrote to Einstein in early January 1915:

\begin{quote}
``I felt the need for a more general theory, as I later attempted to develop and as has actually been advanced by you (and to a lesser extent by Poincar\'{e}) \cite{EinCorr8}.''
\end{quote}

\begin{itemize}
\item[\textbf{Q4:}] \textbf{When does Lorentz tell the truth? In the first textual evidence or in the second?}
\end{itemize}

Lorentz's attitude is at least equivocal but can be simply explained not by textual evidence of course but by the facts.\\

In 1906, Lorentz wrote to Poincar\'{e} to thank him for having sent him his article entitled ``Sur la dynamique de l'\'{e}lectron'' (On the dynamics of the electron) \cite{Poin1906}, all the results of which he had learned from Poincar\'{e}'s letters to him in May 1905. Lorentz was preparing to leave for New York, where he had been invited to give a series of lectures. On the 10$^{th}$ March 1906, Lorentz left Rotterdam for New York, arriving on the 21$^{st}$ March and between this arrival and the 2$^{nd}$ May 1906, he gave a series of lectures at Columbia University in New York on his theory of electrons, which was published in book form in 1909. On page 223 of this book, Lorentz presents Einstein's theory as follows:

\begin{quote}
``The denominations ``effective coordinates'', ``effective time'' etc. of which we have availed ourselves for the sake of facilitating our mode of expression, have prepared us for a very interesting interpretation of the above results, for which we are indebted to Einstein \cite{Lorentz1909}.''
\end{quote}

A few pages later, Lorentz presents his transformation, completed and modified by Poincar\'{e}, with reference to Einstein. He then concludes as follows:

\begin{quote}
``This is a point which Einstein has laid particular stress on, in a theory in which he starts from what he calls the principle of relativity, i.e. the principle that the equations by means of which physical phenomena may be described are not altered in form when we change the axes of coordinates for others having a uniform motion of translation relatively to the original system. I cannot speak here of the many highly interesting applications which Einstein has made of this principle (\ldots)
\end{quote}

As written in the preface, this book has been published in January 1909.

\begin{itemize}
\item[\textbf{Q5:}] \textbf{Given that Lorentz knew all of Poincar\'{e}'s results (including the transformation Poincar\'{e} had named after him), how could he cite Einstein for this transformation and give him all the credit for his discovery?}
\end{itemize}

An attempt of explanation can be found in the race to the Nobel Prize of Physics. Poincar\'{e}'s premature death on the 17$^{th}$ July, 1912 eliminated any possibility of obtaining the Nobel Prize, the regulations prohibiting its award posthumously. In 1912, we discover that Wien nominated Lorentz and Einstein for their ``common'' discovery of the \textit{special theory of relativity}.\\

All of these are facts and not interpretations based on textual evidence.

\subsection{Against Ginoux's ``Fairy Tale'' Claim of Weinstein's comment arXiv: 2510.03793}

In this section, Mrs. Weinstein recalls that I reject the ``oft-repeated claim that Einstein stood outside the scholarly network until Planck ``discovered'' him after 1905''. Then, she adds:

\begin{quote}
Moreover, the claim that Einstein in 1905 lacked direct access to Poincar\'{e}'s May-June writings or to the inner correspondence networks of European physicists is not a ``pure fiction'' but the consensus of documentary historiography [CPAE1, CPAE2, CPAE5]. Historiography, like science, demands clarity.''
\end{quote}

It is well-known that Einstein published eight articles in \textit{Annalen der Physik} between 1901 and 1905. The Editor in Chief of this journal was R\"{o}ntgen as recalled above. Max Planck was then belonging to what was called the \textit{curatorium}. So, could Mrs. Weinstein answer to the following question and all others:

\begin{itemize}
\item[\textbf{Q6:}] \textbf{How is it possible that neither the editor-in-chief (R\"{o}ntgen) nor any of the members of the curatorium (Planck) contacted Einstein about possible corrections to his eight articles before 1905?}
\end{itemize}

At least, the Editor in Chief (R\"{o}ntgen) or one of his associate editors (Planck) should have acknowledged receipt of Einstein's article. But we have no textual evidence of that. From that Mrs. Weinstein concludes that there have been no contact between Einstein and the academics during this period. This is simply impossible! Moreover, according to Max von Laue, then Planck's assistant:

\begin{quote}
``Another area of research in the second half of his career was the theory of special relativity. Planck was one of the first to recognize its importance; at an \textit{unforgettable colloquium}, the first of the 1905/06 winter semester, he presented a paper on Einstein's recently published ``On the electrodynamics of moving bodies''. He corrected an error of reasoning in this famous article, concerning the dynamics of a point mass. It was only then that the same dynamics appeared that H.A. Lorentz had deduced in 1904 from a set of ideas in preparation for the theory of relativity, but which was not identical to it \cite{Laue1947}.''
\end{quote}

Thus, it appears that as early as November 1905, Planck organized a conference on Einstein's work published in the journal in which he was associate editor.

\begin{itemize}
\item[\textbf{Q7:}] \textbf{How is it possible that Planck has not contacted Einstein before this colloquium?}
\end{itemize}

\newpage

\subsection{The Heuristic Method? of Weinstein's comment arXiv: 2510.03793}

In this section Mrs. Weinstein recalls:

\begin{quote}
``For Einstein, ``heuristic'' designated a methodological device-a guiding principle for deriving new results within a
framework that already dispensed with the ether. For Poincar\'{e}, by contrast, the ether was not a heuristic guide at all but an ontological commitment, a
substantive entity posited as the medium of electromagnetic phenomena. To conflate these two uses of ``heuristic'' is to confuse method with substance.''
\end{quote}

After the \textit{textual evidence}, the interpretation of thoughts. I note that Mrs. Weinstein is also a mentalist and that she is able to read the thoughts of the dead.

\subsection{Who Stole the Lorentz Transformation? of Weinstein's comment arXiv: 2510.03793}

In this section Mrs. Weinstein recalls:

\begin{quote}
``The irony is that the first person to publicly refer to them as ``Lorentz transformations'' was Poincar\'{e} himself, in his \textit{Comptes rendus} note of June 1905. Einstein submitted his paper on June 30, 1905. For Einstein to have deliberately avoided the term, he would have needed access to Poincar\'{e}'s unpublished draft--or advance notice from Poincar\'{e}, Lorentz, or the \textit{Comptes rendus} editorial board that Poincar\'{e} was about to use it.''
\end{quote}

This statement demonstrates a complete misunderstanding of the facts. In the letters discovered by Pr. Miller in 1976, we learn that Lorentz asked Poincar\'{e} to help him solve the problem of his transformation, which was not symmetric at the time. Poincar\'{e} succeeded, as Pr. Miller established long before me. Poincar\'{e} then decided to name this transformation after himself, since it was Lorentz who had written it (even incomplete) first, and it was this same Lorentz who asked him to solve this problem. Poincar\'{e} then decided to pay homage to Lorentz, as had been the case years before for Fuchs and Klein. There is no irony in this. It is ridiculous.\\

Moreover, let us notice another problem in the analysis of the chronology of events. Many historians of science consider that the manuscript submitted by Einstein on June 30$^{th}$ entitled ``Zur Elektrodynamik bewegter K\"{o}rper,'' (On the electrodynamics of moving bodies) \cite{Einstein1905} is exactly the same as the version published of Einstein on September 26, 1905.

\begin{itemize}
\item[\textbf{Q8:}] \textbf{How can they be sure of that?}
\end{itemize}

How can they be sure of that? Thanks to the \textit{textual evidence} would certainly say Mrs. Weinstein. To be sure of that, you just need to compare the original manuscript written by Einstein in May-June 1905 and sent to the Editor in Chief of \textit{Annalen der Physik}. The process of submission of an article is still the same for ages. You write an article that you send to an Editor in Chief or to an associate editor. This latter send it to one or several reviewers (at least two in general) which send back to the Editor in Chief or to the associate editor a recommendation. The reviewers  propose either to accept the paper in its state, to reject it or to ask for corrections or modifications. Unfortunately, there is another problem here. In November 1943, Einstein, who had been a refugee in the United States for ten years, decided to sell the manuscript of his 1905 article to participate in the war effort. According to the \textit{New York Times} of February 2, 1944:

\begin{quote}
``The relativity manuscript was especially re-copied by Dr Einstein for this occasion to replace the original manuscript which he threw away after its
publication in 1905.''
\end{quote}

Under the dictation of his secretary, Helen Dukas, Einstein therefore rewrote the article as it had been published in \textit{Annalen der Physik}. It is therefore impossible to highlight any similarity between the original manuscript sent by Einstein on June 30 and the version published on September 26, 1905. No textual evidence can be used here. But, Mrs. Weinstein will certainly explain to us that it must be the same article and that there was no difference between the two versions (handwritten and published) since it is Einstein's. I'll leave it to the reader to appreciate the academic rigor Mrs. Weinstein prides herself on, and the strength of the arguments she uses to convince us at all costs of their veracity.

\subsection{Same But Different of Weinstein's comment arXiv: 2510.03793}

In this section Mrs. Weinstein recalls:

\begin{quote}
``Ginoux's charge of ``falsehood'' rests on a narrow reading of ``La mesure du temps.'' He notes, triumphantly, that the words ``luminiferous ether'' do not
appear in that text. But absence of a phrase is not absence of a framework.''
\end{quote}

Yes, I agree. But the same holds for Einstein. In his famous article, Einstein \cite{Einstein1905} studies the propagation of a light signal from one frame of reference to another. At that time, Lorentz, Planck, Poincar\'{e} and nearly all the scientific community considered that light propagates in the \textit{luminiferous ether}. Einstein says nothing about this medium except that it is \textit{superfluous}, i.e. that he doesn't need it for his demonstration. And this is really the case. But, in 1920 before Lorentz, he declared:

\begin{quote}
``Recapitulating, we may say that according to the \textit{general theory of relativity} space is endowed with physical qualities; in this sense, therefore, \textbf{\textit{there exists an ether}}. According to the \textit{general theory of relativity} \textbf{\textit{space without ether is unthinkable}}; for in such space there not only would be no propagation of light, but also no possibility of existence for standards of space and time (measuring-rods and clocks), nor therefore any space-time intervals in the physical sense. But this \textbf{\textit{ether}} may not be thought of as endowed with the quality characteristic of ponderable media, as consisting of parts which may be tracked through time. The idea of motion may not be applied to it \cite{Einstein1921}.''
\end{quote}

So, according to Mrs. Weinstein, if we admit that Einstein always tells the truth, many problems arise here. First, Einstein reintroduces the ether that he had supposedly banished. Then, he explains that ``space without ether is unthinkable; for in such space there not only would be no propagation of light.'' But in 1905, in what medium was propagating light in his article? Depending on who is using the \textit{luminiferous ether}, it becomes for Mrs. Weinstein ``the decisive reconfiguration that gave birth to special relativity.''

\subsection{Einstein's Method vs. Ginoux's Historiography of Weinstein's comment arXiv: 2510.03793}

In this section, Mrs. Weinstein recalls Einstein's full quote.

\begin{quote}
``The question as to the independence of those relations is a natural one because the Lorentz transformation, the real basis of the special relativity theory, in itself has nothing to do with the Maxwell theory and because we do not know the extent to which the energy concepts
of the Maxwell theory can be maintained in the face of the data of molecular physics \cite{Einstein1935}.''
\end{quote}

Yes, this is true. But then, why did Einstein apply the Lorentz transformation to the Maxwell equations in his 1905 article? Here we see the limits and inconsistency of Ms. Weinstein's over-used \textit{textual evidence}. This quote also highlights Einstein's ability to rewrite the history of the development of the theory of special relativity.

\subsection{The Title of Einstein's Paper of Weinstein's comment arXiv: 2510.03793}

In this section Mrs. Weinstein recalls that I have shown that the entire title of Einstein's article \cite{Einstein1905} was fully contained in one of the sentences of Poincar\'{e}'s concluding remarks at Saint Louis:

\begin{quote}
``Shouldn't we also strive for a more satisfactory theory of the electrodynamics of moving bodies? \cite{Poin1904}''
\end{quote}

To this argument of \textit{textual evidence}, Mrs. Weinstein replies that it's pure coincidence! How marvelous. She has based almost all her arguments on \textit{textual evidence} and, in the face of this obvious evidence, she considers that it is not \textit{textual evidence}. Once again, her textual evidence is Einstein-dependent or variable-geometry.

\subsection{Heuristic Freedom, Not Foreknowledge of Weinstein's comment arXiv: 2510.03793}

Here Mrs. Weinstein explains that the factor $\varphi(v) = a(v)\gamma$ is a ``mathematical convenience'' taken by Einstein to state the transformation established at least three weeks before by Poincar\'{e}. I can just repeat the words of Pr. Arthur I. Miller:

\begin{quote}
``How did Einstein know that he had to make the further substitution $a = \varphi(v)\sqrt{ 1 - v^2/V^2 }$ in order to arrive at those space and time transformations in agreement with the postulates of relativity theory? \ldots But why did Einstein make this replacement. It seems as if he knew beforehand the correct form of the set of relativistic transformations \cite{Miller1981}.''
\end{quote}

\subsection{On the Relativity of Irony of Weinstein's comment arXiv: 2510.03793}

In this section, Mrs. Weinstein recalls with a certain irony the preface of my book by Pr. Arthur Miller which explains that ``Einstein has discovered the special theory of relativity and not Poincar\'{e}''. This is very surprising. It suggests that Mrs. Weinstein has not read my book to the end. If that were the case, she would have been able to see that I never wrote or claimed that Poincar\'{e} discovered the theory of special relativity. Quite the contrary. In my book I have concluded by these words:

\begin{quote}
So it's hard to understand why, for over seventy years, many scientists and historians of science have stubbornly tried to make \textit{Poincar\'{e} the father of relativity, in spite of himself} \cite{Ginoux2024}.
\end{quote}

This \textit{textual evidence} is crystal clear but may be not for Mrs. Weinstein, this being said without any irony.

\subsection{A Long and Tedious Computation of Weinstein's comment arXiv: 2510.03793}

In this section, Mrs. Weinstein recalls that I have said that Einstein has derived the relativistic addition law after a long and tedious computation. Weinstein explains that my presentation, ``unfortunately, rests on a mistaken premise of the mathematics at issue''. The question here is not whether Ginoux did not understand the mathematical premises at issue but whether Einstein's demonstration is indeed long and tedious or not. The simple comparison between Poincar\'{e}'s and Einstein's demonstration enables to show that Poincar\'{e}'s demonstration \cite{Poin1906} takes two lines and Einstein's at least one page \cite{Einstein1905}.

\subsection{Einstein's Citation Practice of Weinstein's comment arXiv: 2510.03793}

In this section, Mrs. Weinstein comes back on the fact that I pointed out Einstein's absence of citations in his article of June 1905. She explains that:

\begin{quote}
``To suggest that Einstein's silence was a strategy of suppression is not historiography but innuendo. The historian's task is not to treat absence as guilt but to test claims against the surviving documentary record.''
\end{quote}

The problem with Einstein is that he has tried to make disappear a lot of embarrassing documents. Just two examples are enough to prove that. The first is the discovery of the famous Love Letters between Einstein and his first wife Mileva Maric. These letters discovered by Einstein's granddaughter Evelyn were published in 1992. They revealed that Einstein and Mileva Maric had a daughter named Lieserl born in 1902, of whom we have lost all trace. The second is a strange letter discovered in 1999, dated May 25, 1918, which Ilse Einstein (one of the daughters of his second wife Elsa Einstein) sent to her close friend Georg Nicolai, doctor, famous pacifist activist and friend of the Einstein family, with the words:

\begin{quote}
``Destroy this letter immediately after reading it,''
\end{quote}

scrawled across the top in big letters. In this letter, Ilse explained to Nicolai that one afternoon had suddenly escalated into a serious proposal that Einstein
marry her instead of her mother. Einstein, she said, had confessed that he loved her. Certainly, Mrs. Weinstein will reply (as she has already done in her comment) that my ``reasoning collapses private life into a wholesale claim of intellectual dishonesty''. The discovery of these documents sheds new light on Einstein's behavior and his ability to conceal facts embarrassing to himself. The question, as Mrs. Weinstein points out, is whether one can dissociate the man from his work. I believe it is impossible to dissociate the man from his work, and I cannot be satisfied with Mrs. Weinstein's romantic, not to say romanticized, explanation of the absence of quotations in Einstein's June 1905 article.

\subsection{Einstein's 1920 ``Ether'' Was Not Lorentz's Ether of Weinstein's comment arXiv: 2510.03793}

In my previous comment to Mrs. Weinstein's review of my book (arXiv:2509.09361), I have noticed that Einstein had considered \textit{luminiferous ether} as superfluous (\"{u}berflu\"{u}ssig in German) in his June 1905 article and resurrected it in 1920 in a lecture before Lorentz. To this, Mrs. Weinstein first give me a lesson of etymology by explaining that \"{u}berfl\"{u}ssig means obsolete in German. Although, Ich habe Deutsch gelernt (I learned German), I do not pretend to be a specialist of this language. Nevertheless, when I have asked to native German speakers if the meanings of \"{u}berfl\"{u}ssig was obsolete, they disagreed. But this is not the problem here. The question is to understand why Einstein did that. Mrs. Weinstein explains then:

\begin{quote}
``Lorentz still clung to the ether, and Einstein, who revered him, chose to frame his redefinition of the metric field of general relativity in
language Lorentz could recognize.''
\end{quote}

Thus, from the \textit{textual evidence}, Mrs. Weinstein explains to us that Einstein is lying before Lorentz because he ``revered him''.

\begin{quote}
``As Stachel has noted, the ether that Einstein reintroduced ``differed fundamentally from the ether he had banished'' [Sta-01].
To ignore this discontinuity is to replace conceptual analysis with a slogan.''
\end{quote}

Here, Mrs. Weinstein repeats the same sentence of John Stachel like an advertising slogan and even has the audacity to reproach me ``to replace conceptual analysis with a slogan.''

\subsection{Einstein's Light Postulate: An Indispensable Principle, Not a Curiosity of Weinstein's comment arXiv: 2510.03793}

In this section Mrs. Weinstein recalls that I have said that ``I still claim that Einstein's 1905 light postulate is curious, since in my opinion it should have been like for Poincar\'{e} a consequence of the \textit{relativity principle} and not the contrary.'' Contrary to what Mrs. Weinstein claims, I have never considered the invariance of light to be a curious postulate. I said it was curious that Einstein should consider this invariance incompatible with the principle of relativity. Einstein wrote in his famous article:

\begin{quote}
``We will raise this conjecture (the purport of which will hereafter be called the ``Principle of Relativity'') to the status of a postulate, and also introduce another postulate, which is only apparently \textbf{irreconcilable} with the former, namely, that light is always propagated in empty space with a definite velocity $c$ which is independent of the state of motion of the emitting body \cite{Einstein1905}.''
\end{quote}

In my opinion, the invariance of velocity of light is not a postulate \textbf{irreconcilable} with the \textit{principle of relativity} but a consequence of this principle. However, I am pretty sure that Mrs. Weinstein will reply by saying with a certain irony that I didn't understand Einstein.

\subsection{Algebraic Form Is Not Physical Content of Weinstein's comment arXiv: 2510.03793}

Here, Mrs. Weinstein considers that Poincar\'{e}'s formula for the relativistic velocity addition law is not a relativistic velocity addition law since it is written in an algebraic form based on group theory. This is simply ridiculous! Nearly all formulas in Physics are written in an algebraic form and most of the time in a dimensionless algebraic form. According to Mrs. Weinstein, we should thus consider that all these formulas or laws are not belonging to a Physical content. Moreover, in Einstein's article of June 1905, he wrote in section \S 5 after his long and tedious demonstration for establishing the relativistic velocity addition law:

\begin{quote}
``we see that such parallel transformations-necessarily-form a group \cite{Einstein1905}.''
\end{quote}

This sentence is written without any demonstration. This is also quite surprising since in neither of his previous publications, i.e. between 1901 and 1905, Einstein has never spoken about group theory. On the contrary, as soon as 1880, Poincar\'{e} was already deeply involved in group theory notably with Felix Klein.

\begin{itemize}
\item[\textbf{Q9:}] \textbf{How Einstein did he know without any demonstration that this transformation forms necessarily a group?}
\end{itemize}

\subsection{On the Alleged ``Coincidence'' with Lorentz and Poincar\'{e} of Weinstein's comment arXiv: 2510.03793}

In this section Mrs. Weinstein recalls that I have made the two following assertions that I still maintain:

\begin{quote}
``1. Lorentz's 1904 expressions for velocity and charge density were wrong and,\\
2. Einstein's results are ``the same as those obtained in May-June 1905 by Poincar\'{e}.''''
\end{quote}

Concerning point 1., in her answer, Mrs. Weinstein explains that Lorentz's expressions for the velocity and charge density are ``not wrong at all.'' Her attitude is truly astonishing. In several of her paragraphs, Mrs. Weinstein explains the fundamental importance of \textit{textual evidence}, and when one is presented to her, she pretends to ignore it. After Poincar\'{e}'s untimely death on the 17$^{th}$ July, 1912, Lorentz wrote in 1913 an analysis entitled ``Deux m\'{e}moires de Henri Poincar\'{e} sur la physique math\'{e}matique,'' (Two Papers of Henri Poincar\'{e} on Mathematical Physics) only published in 1921 because of the First World War. Lorentz then recognized:

\begin{quote}
``Poincar\'{e}, on the contrary, obtained a perfect invariance of the equations of electrodynamics, and he formulated the \textit{postulate of relativity},
terms which he was the first to employ. Indeed, stating from the point of view that I had missed, he found the formula ([of the velocity and charge density]). Let us add that by \textbf{correcting} the imperfections of my work he never reproached me for them \cite{Lorentz1921}.''
\end{quote}

Thus, Lorentz was the first to recognize that his expressions for velocity and charge density were wrong. How can Mrs. Weinstein so deny the \textit{textual evidence} she uses in all her arguments?

\begin{itemize}
\item[\textbf{Q10:}] \textbf{If Lorentz's expressions for the velocity and charge density ``were not wrong at all'', why did Lorentz recognized that they were indeed?}
\end{itemize}

Moreover, Pr. Arthur I. Miller, who is a well-known specialist of the theory of special relativity, wrote fifty years ago:

\begin{quote}
``The field transformation equations are correct, but the ones for velocity and charge density are not (\ldots)\\
Removing this asymmetry enabled Poincar\'{e} to derive the correct set of transformation equations for the velocity and charge density \cite{Miller1973}.''
\end{quote}

But Mrs. Weinstein certainly has a better understanding of the issue than this great specialist.

Concerning point 2., Mrs. Weinstein explains:

\begin{quote}
``In his \textit{Rendiconti} memoir, Poincar\'{e} did obtain the correct velocity transformation formulas. Still, by a different route \ldots ''
\end{quote}

Her argument is still: this is the equations are the same but they are different. Incredible!

\subsection{On the Originality of My Reconstructions of Weinstein's comment arXiv: 2510.03793}

In this section, Mrs. Weinstein recalls that I have said that ``she had only recopied some results already published in my book and in the contributions of Pr. Miller''. As a consequence, Mrs. Weinstein considers that I accuse her of plagiarism. Although I stand by my assertion, I never considered that she had plagiarized my work or others and I never questioned her scholarly integrity and for one obvious reason: the mathematical demonstrations of Poincar\'{e} and Einstein are in the public domain. What I have asserted, and what I continue to assert, is that there is no originality in her reconstruction or in mine, since the mathematical demonstrations of Poincar\'{e} and Einstein have long since been thoroughly analyzed by great scientists and historians of science such as Pr. Miller \cite{Miller1973, Miller1980, Miller1981}. \newpage

Here I could paraphrase Mrs. Weinstein by saying that it is the same but it is different. What is different is her interpretation of the results for which I do not agree. But equations and demonstrations are exactly the same. An example is presented in Fig. \ref{fig5} below.

\begin{figure}[htbp]
\includegraphics[width=16cm,height=18cm]{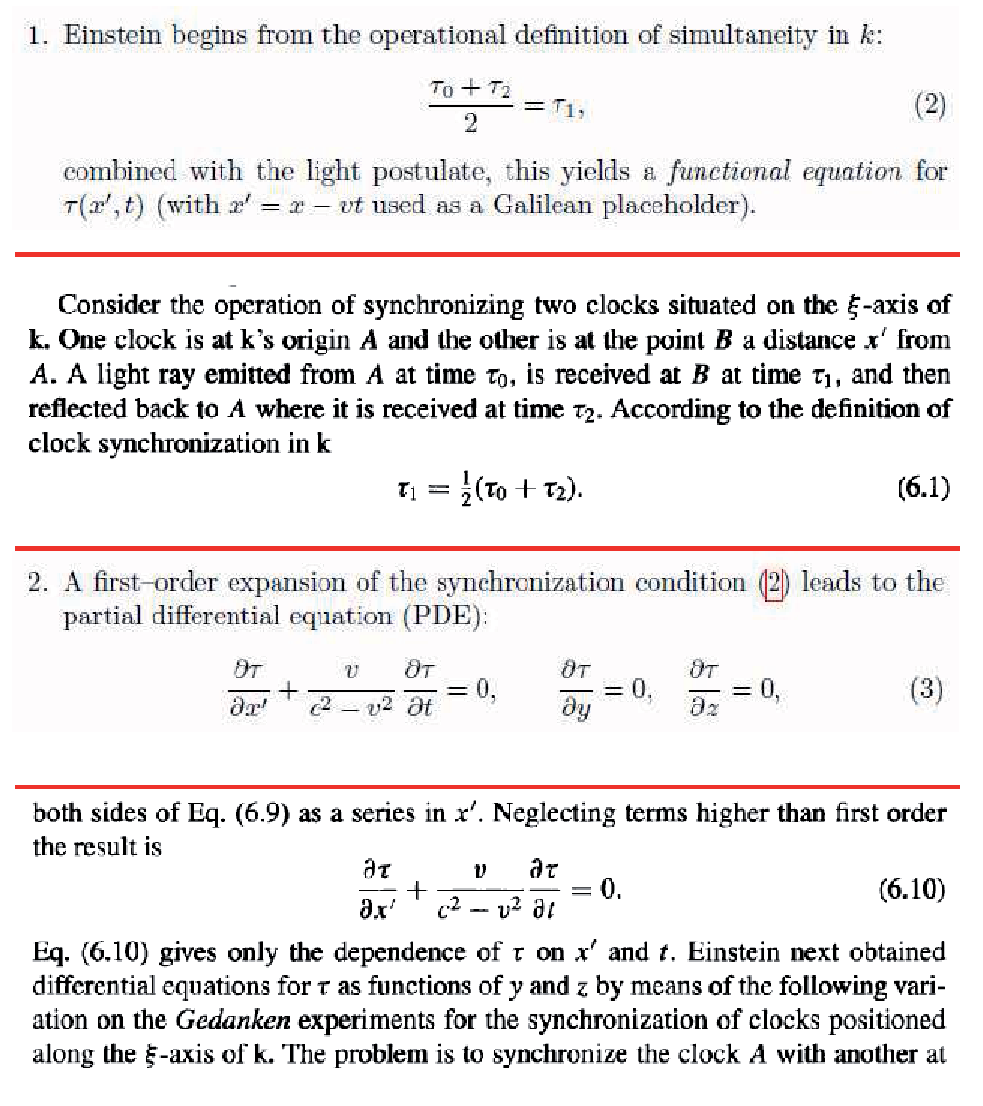}
\caption{Screen shot of Weinstein's first comment and chapter 6 of Miller \cite{Miller1981}}
\label{fig5}
\end{figure}

Equations (2) and (3) of Fig. \ref{fig5} are Weinstein's reconstruction. Equations (6.1) and (6.10) are Miller's reconstruction \cite{Miller1981} presented in the chapter 6 of his book. As one can observe there are no difference between Weinstein's and Miller's equations. And I could continue thus line-by line comparison between Mrs. Weinstein's comment and Pr. Miller's chapter 6 of his book \cite{Miller1981}. So, where is the novelty? In the interpretation? Of course, but a credible interpretation needs solid arguments. This is clearly not the case for Mrs. Weinstein. But a credible interpretation needs solid arguments. This is clearly not the case for Weinstein. In this regard, I am sad to see that Mrs. Weinstein, having no more arguments to oppose me, is reduced to insulting me by quoting a letter from Einstein to Grossmann and comparing me with a ``donkey''. So, she explains in her commentary that it is essential to have respectful conduct among academics, she provides us here with proof of her abilities in this area.

\subsection{Historiography Is Not Rhetorical Characterization of Weinstein's comment arXiv: 2510.03793}

In this section, Mrs. Weinstein explains that:

\begin{quote}
``That principle is precisely what I have upheld throughout my analysis: adherence to the extant documentary record, refusal to speculate on lost or hypothetical
letters, and attention to the conceptual architecture of theories rather than to coincidences of terminology or algebra. To insist, without evidence, that Einstein concealed knowledge of Poincar\'{e}, or that citation practices prove suppression, is not to ``submit to facts,'' but to subordinate facts to suspicion.''
\end{quote}

If only Mrs. Weinstein had done everything she said. Unfortunately, it cannot be said that this was not the case, as has been demonstrated in this document.

\subsection{A Final Word, and It Is Einstein's of Weinstein's comment arXiv: 2510.03793}

In her last section, Mrs. Weinstein points out with irony that in his very last lecture in July 1912, Poincar\'{e} mentioned Einstein's June article of 1905. This \textit{textual evidence} proves on the one hand that he had probably read this paper and highlight on the other hand the main difference between these two scientists: Poincar\'{e} was very careful to cite his sources and Einstein was not.

\subsection{Conclusion}

Contrary to what she pretends, Mrs. Weinstein's arguments are essentially based on what she calls \textit{textual evidence} which depend on reader's interpretations. But, these interpretations can lead to one thing and its opposite, and therefore cannot be considered as formal proof. According to Leibniz:

\begin{quote}
``Descartes lodged the truth in the hostel of evidence but he forgot to give us the address.''
\end{quote}

The same holds for Mrs.  Weinstein.\\

Mrs. Weinstein's analysis is clearly based on a desire to defend Einstein at all costs. It is therefore biased and subjective. Her arguments have no value because they are not based on documents, archives, letters, etc., but on her interpretations of these documents, archives, and letters. She is never able to prove anything she claims. She tries desperately to defend her point of view with arguments that are irrelievable. She uses \textit{textual evidence} as an irrefutable argument, while it ultimately turns against her as I have demonstrated in this document. Finally, when she has no more arguments to oppose me, she chooses condescension and insults by explaining that my ``presentation, unfortunately, rests on a mistaken premise of the mathematics at issue'' and that I am a donkey. She even has the audacity to lecture me and explain that we must have a respectful attitude among academics. Where is the respect in her comment?\\

\end{document}